\documentclass[preprint,12pt]{article}

\usepackage{graphicx}
\usepackage{epstopdf}
\usepackage{subfigure}
\usepackage{array}
\usepackage{multirow}
\usepackage{slashbox}
\usepackage{pdflscape}
\usepackage{rotating}
\usepackage{geometry}
\usepackage{pdflscape}
\usepackage{caption}
\usepackage{color}
\usepackage{authblk}

\usepackage{amssymb}

\title{Explosive Percolation: Unusual Transitions of a Simple Model}

\author[1]{N. Bastas}
\author[1]{P. Giazitzidis}
\author[1,2]{M. Maragakis}
\author[1,3,4]{K. Kosmidis}

\affil[1] {Physics Department, Aristotle University of Thessaloniki,Thessaloniki 54124,Greece}
\affil[2] {Department of Economics, University of Macedonia, Thessaloniki, Greece }
\affil[3] {Computing \& Communications Center, Aristotle University of Thessaloniki,Thessaloniki 54124,Greece}
\affil[4] {School of Engineering and Science,Jacobs University Bremen, Campus Ring 1, 28759 Bremen, Germany.}

\begin{document}

\maketitle



\begin{abstract}
In this paper we review the recent advances on explosive percolation, a very sharp phase transition first observed by Achlioptas et al. (Science, 2009). There a simple model was proposed, which changed slightly the classical percolation process so that the emergence of the spanning cluster is delayed. This slight modification turns out to have a great impact on the percolation phase transition. The resulting transition is so sharp that it was termed explosive, and it was at first considered to be discontinuous. This surprising fact stimulated considerable interest in ``Achlioptas processes''.  Later work, however, showed that the transition is continuous (at least for Achlioptas processes on Erd\"{o}s networks), but with very unusual finite size scaling. We present a review of the field, indicate open ``problems'' and propose directions for future research.

\end{abstract}




\section{Introduction}

Percolation is a very important topic in the Statistical Physics of phase transitions \cite{Bunde_book,Stauffer_Book}. Although its first appearance in the scientific literature date back to the year 1940 and the work of Flory \cite{Stauffer_Book} it remains a challenging and active field of research even today. A brief search in a scientific search-engine like ``Scopus'' reveals that last year (2012) more than 1800 scientific papers containing the word ``Percolation'' in their abstracts have been published in peer-reviewed journals. This is a clear indication of the activity of the field and its interest for the scientific community. Percolation represents a paradigmatic model of a geometric phase transition. In the classical site percolation model, the sites of  a square lattice are randomly occupied with particles with probability $p$, or remain empty with probability $1-p$. Neighboring occupied sites are considered to belong to the same cluster. Percolation theory simply deals with the number and properties of these clusters. When the occupation probability $p$ is small, the occupied sites are either isolated or form very small clusters. On the other hand, for large $p$ there are a lot of occupied sites that have formed one large cluster. It is in fact possible to find several paths of occupied sites which a walker can use to move from one side of the lattice to the other. In this latter case, it is said that a giant component of connected sites exists in the lattice. This ``infinite cluster'' as it is called does not appear in a gradual ``linear'' way with increasing $p$. It appears above a critical occupation probability $p_{c}$. Below $p_c$ there are only small clusters and even if we increase the lattice size considerably, these clusters remain small, i.e. the size of the largest cluster does not depend on the system size. Above $p_c$, suddenly, small clusters join together to form a single large cluster whose size scales with system size. Hence, the term giant component or infinite cluster which is very common in the literature \cite{Bunde_book}. 

Thus, the phase transition related to Percolation is a geometric one, i.e. the appearance of an ``infinite'' connected cluster. In Physics, phase transitions are usually thermally induced, meaning that a property appears above or below a characteristic temperature. During a phase transition of a given medium certain properties of the medium change, often discontinuously, as a result of some external condition, such as temperature, pressure, and others. For example, a liquid may become gas when heated up to the boiling point, resulting in an abrupt change in volume. Phase transitions can be described by determining the behavior of an ``order parameter''. The order parameter is normally a quantity which is zero in one phase (usually above the critical point), and non-zero in the other. For a liquid-gas transition the difference of the densities of the gas $\rho_{gas} $ and the liquid phase $\rho_{liq}$, $ \rho=|\rho_{liq}-\rho_{gas}|$ is an appropriate order parameter.

Phase transitions are marked by a singularity of the free energy or one of its derivatives at the transition point.
First-order phase transitions exhibit a discontinuity in the first derivative of the free energy with respect to some thermodynamic variable.
In this case the order parameter is discontinuous, as for example in a solid-liquid phase transition. 
First-order phase transitions are those that involve a latent heat. During such a transition, a system either absorbs or releases a fixed (and typically large) amount of energy. During this process, the temperature of the system will stay constant as heat is added: the system is in a "mixed-phase regime" in which some parts of the system have completed the transition and others have not. Familiar examples are the melting of ice or the boiling of water (the water does not instantly turn into vapor, but forms a turbulent mixture of liquid water and vapor bubbles).

When the change of the order parameter is not discontinuous, we usually talk about second-order transitions. Second-order phase transitions are also called continuous phase transitions. They are characterized by a divergent susceptibility, an infinite correlation length, and a power-law decay of correlations near criticality. Examples of second-order phase transitions are the ferromagnetic transition, the superconducting transition and the superfluid transition.

 In percolation, $p$ plays the same role as the temperature in thermal phase transitions, i.e. that of the control parameter, while the order parameter  is the probability $P_{\infty}$ that a site belongs to the infinite cluster. Classical percolation exhibits all the characteristics of a continuous phase transition.
 For $p>p_c$, $P_{\infty}$ increases with $p$ by a power law
 \begin{equation}
 P_{\infty} \sim (p-p_c)^{\beta}
 \label{eq1}
 \end{equation}  
 Other important quantities are the correlation length $\xi$ which is defined as the mean distance between two sites on the same finite cluster and the mean number of sites $S$ of a finite cluster. When $p$ approaches $p_c$, $\xi$ increases as 
 \begin{equation}
 \xi \sim (p-p_c)^{-\nu}
 \label{eq2}
 \end{equation}
 The mean number of sites $S$ of a finite cluster also diverges at $p_c$
 \begin{equation}
 S \sim (p-p_c)^{-\gamma}
 \label{eq3}
 \end{equation} 
 The critical exponents $\beta,\nu$ and $\gamma$ describe the critical behavior associated with the percolation transition and are universal. Although, the percolation threshold changes for even slight modifications of the model (for example, site and bond $p_{c}$'s are different) the critical exponents are very robust in changing the percolation model details \cite{Schrenk2013b}. They do not depend on the structure of the lattice (e.g., square or triangular) or on the type of percolation (site, bond or even continuum) \cite{Stauffer_Book}.
 
Thus, there was a big surprise when recently, Achlioptas et al \cite{Achlioptas2009a}, claimed that a rather simple modification of the original percolation process 
leads to a new seemingly first-order percolation transition. This transition was named ``Explosive percolation''(EP) due to the abrupt character of the transition. However, percolation models which are characterized by first-order transition where reported more than $20$ years ago. For example, in \cite{Adler1991}, bootstrap percolation on hypercubic lattices was shown to exhibit discontinuous transition for certain bounds. What is new with this process is that it depends on the site (or bond) occupation history, thus falling into the category of non-equilibrium processes (unlike other percolation models). In this paper, we review the recent developments on the investigation of this ``explosive'' transition presenting the most important results concerning the nature of the transition as well as the several variants of the original model.

Before proceeding to the main text, we must emphasize that we restrict ourselves in the case of single layered networks which consist of undirected ``connectivity'' links. Recent studies on recursive cascading failures, extend the concept of first-order transition to other types of systems. For example, it is shown that the presence of ``dependency'' links in a single network induces a critical threshold which seperates the first from the second-order region \cite{Parshani2011}. Moreover, in  the case of interdependent networks, for different interconnection patterns, number of layers and type of links, the systems exhibit first-order transitions \cite{Buldyrev2010,Parshani2010,Gao2011}. All the previous results are proven analytically and sustained by simulations. The interested reader should refer to the cited articles.


\section{What is Explosive percolation}

Physicists are usually interested in studying the percolation transition in lattices, as such a transition has important applications in solid state physics. Mathematicians on the other hand find it more convenient to study percolation on graphs, especially Erd\"{o}s -R\'{e}nyi (ER) random graphs \cite{Erdos1960}, where exact results can be obtained. Actually, the percolation transition of ER random networks is one of the most studied phenomena in probability
theory. Consider a graph of $N$ initially unconnected nodes and start randomly connecting them. When $rN$ edges have been added, if $r < 1/2$, the largest component remains
small, i.e. its number of vertices $C$ scaling as $log N$. In contrast, if $r > 1/2$, there is a component of size
linear in $N$. Actually, $C \simeq (4r-2)N$ when $r$ is slightly greater than $1/2$ and, thus, the fraction of vertices
in the largest component undergoes a continuous phase transition at $r = 1/2$. The mechanism of percolation on random graphs is rather similar to bond percolation on lattices. 
Of course, the critical exponents are different as it is known that these exponents depend on the dimensionality of the underlying space.
Graphs can usually be considered as infinite dimensional systems. In such a case we say that the two models belong to different universality classes.
One normally expects to find different universality classes in models where the dimensionality of the space or the underlying symmetries are different. 
However, as mentioned above, this is not expected to be observed in small modifications of the model that generates the percolation transition.
In such cases, one observes that the critical point changes depending on the details of the model but the critical exponents do not. This concept is known as ``Universality''.


\subsection{Early Results}

\begin{figure}
\begin{center}
\includegraphics[angle=0,width=12cm]{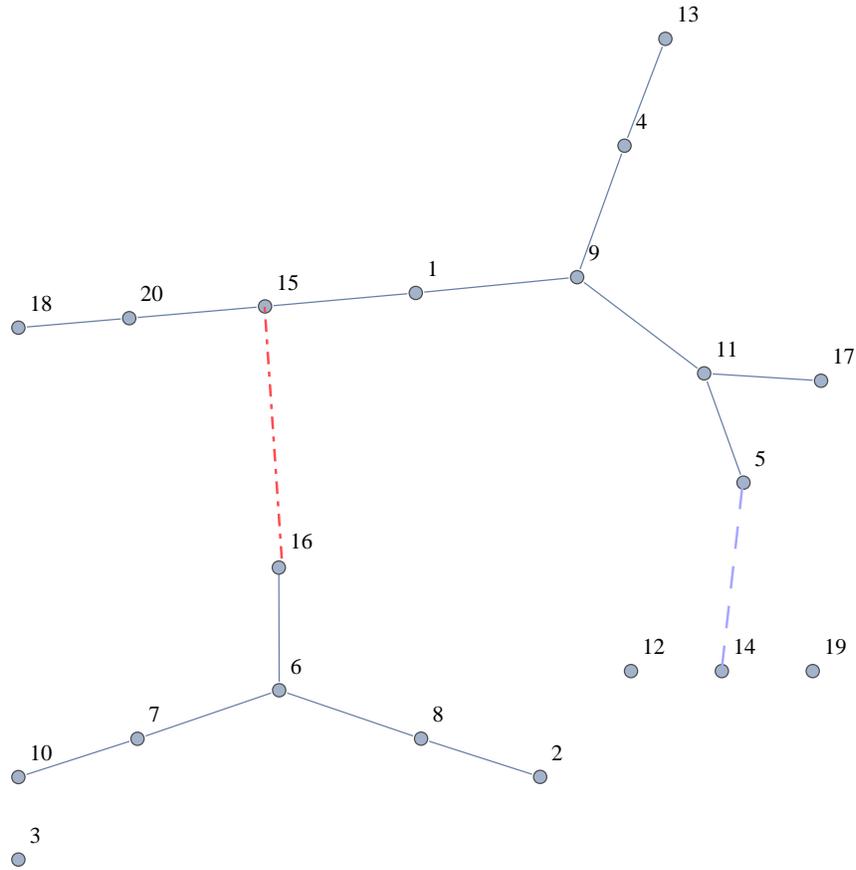}
\end{center}
\caption{(Color online) Schematic of the ``Product rule'' for explosive percolation.Two trial edges, between nodes 15-16 (red dotted line) and between nodes 5-14 (blue dashed line) are initially selected. The edge 5-14 will be kept as the product of the cluster sizes it connects is smaller. The edge 15-16 will be discarded for this trial, but for a future trial, the same edge may be accepted.}
\label{Schema1}
\end{figure}

In \cite{Achlioptas2009a} the authors proposed a new method for the construction of a network (graph)  which produces ``explosive'' transitions: 
We start from a set of $N$ unconnected nodes i.e. initially there are no edges.
To construct an ER graph one would start to randomly insert edges in the graph. Here, when filling sequentially the initially empty network, instead of randomly inserting a bond, we choose two candidates
and investigate which one of them leads to the smaller cluster sizes. The one that does this is kept as a new bond in the network, while the second one is discarded. 
Such a procedure is in clear contrast to the gradual growth of the largest component in the random percolation model. This is because the largest cluster is generally disfavored in growing. Thus, the emergence of a giant component is considerably slowed down and the rate at which it eventually grows is much larger in the end. It is in fact so high that it is considered to form abruptly and thus the behavior of the order parameter was termed ``explosive''.
There are several selection rules to achieve this. The authors propose the ``product rule''. From the two trial bonds inserted in the network they choose the one that minimizes the product of the sizes of the components it merges.
In fig. \ref{Schema1} we present a schematic description of the ``product rule''. At a given instance the network is comprised of a cluster of 10 nodes,
a cluster of 6 nodes and 4 clusters of size one.
Two trial edges are selected. One between nodes 15-16 (red dotted line) and another between nodes 5-14 (blue dashed line). The blue edge will be kept as the product of the cluster sizes that connects ($10 \times 1=10$) is smaller than that of the red edge ($10 \times 6=60$).

\begin{figure}
\begin{center}
\includegraphics[angle=-90,width=12cm]{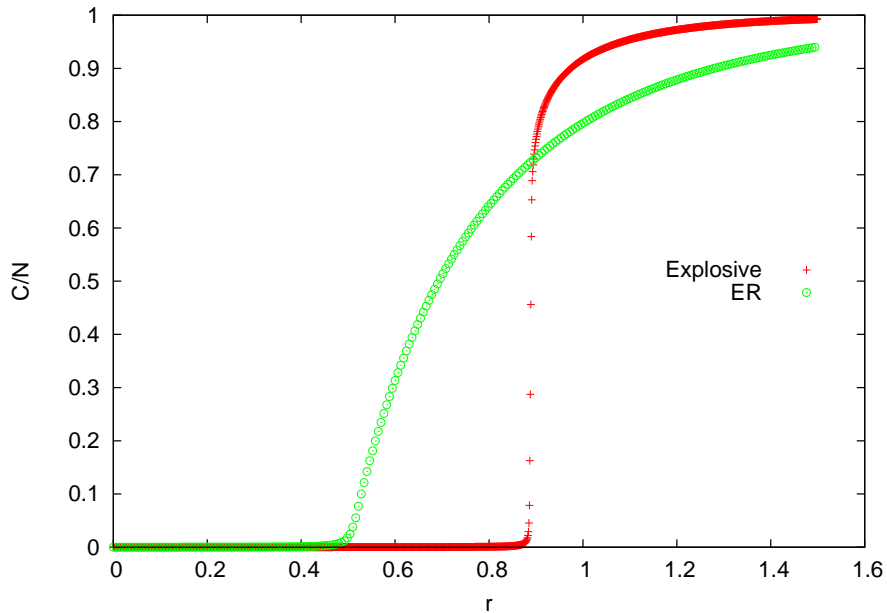}
\end{center}
\caption{(Color online) Fraction of nodes belonging to the largest cluster $C/N$ as a function of the edges fraction $r$ for a network of $N=100000$ nodes. Green circles: ER network, red crosses: Product Rule.The results are averages over $100$ different system realizations.}
\label{fig1}
\end{figure}

Other selection rules, such as the ``sum rule'', lead to qualitatively similar results. 
In Fig.\ref{fig1} we plot the fraction of nodes belonging to the largest cluster $P_{\infty}=C/N$ as a function of the edges fraction $r$ for a network of $N=100000$ nodes. The (green) circles show the classical ER network. The (red) crosses correspond to the ``product rule'' selection. All points are data from Monte Carlo simulations and represent averages over $100$ different system realizations. 

It is immediately obvious that in the case of the ``product rule'' the transition is much more sudden and abrupt than in the classical case. In fact, the change is so abrupt that it looks as a first-order transition with a discontinuous jump of the order parameter at a critical $r_{c}$.

In fact, this is what was initially thought and the main claim of \cite{Achlioptas2009a} was that the ``product rule'' in graphs leads to a first-order percolation transition. This claim, however, was shown to be
false later \cite{Riordan2011a}.

The main problem with the numerical investigations of percolation is that it is very difficult to distinguish between a first-order transition where the order parameter changes discontinuously and a continuous transition with a very small $\beta$ exponent (see Eq.\ref{eq1}). In order to overcome this difficulty the authors in \cite{Achlioptas2009a} proposed the $\Delta$ criterion. Let $C$ denote the size of the largest component,
$t_0$ denote the last step for which $C < N^{1/2}$, and $t_1$ the
first step for which $C > 0.5N$. We define $\Delta=t_1-t_0 $.
The authors in \cite{Achlioptas2009a} claim that in continuous transitions,
the interval $\Delta$ is always extensive,i.e., linear in $N$. For explosive percolation (product rule) they find that $\Delta \sim N^{2/3}$.

The $\Delta$ criterion seems to be rather plausible. In percolation (whether classical or explosive) we have a transition between one phase (non percolative) where the largest cluster does not scale with the system size and another phase where the largest cluster occupies a finite fraction of the system i.e. it is proportional to $N$. $\Delta$ measures how many bonds one should add in order to ``pass'' from one phase to the other. For an explosive transition $\Delta/N$ tends to zero for $N \rightarrow \infty$ while for the classical case it tends to a non-zero value. Since a negligible fraction of bonds has to be added to the system in the explosive case in order to pass
to the percolative phase, the explosive percolation transition was reported to be a first-order transition.

The $\Delta$ criterion is rather charming as it provides a ``simple'' way to distinguish the order of a transition, which is a very difficult task when analyzing Monte Carlo simulation data or other numeric results. Problems, however, became almost immediately apparent when the explosive percolation process was studied on a square lattice instead of a network \cite{Ziff2009,Ziff2010a} .

More specifically, Ziff \cite{Ziff2009} studied the effects of the ``product rule'' when it is used to add bonds in a regular square lattice and compared the results with those of classical bond percolation. Indeed the observed transition was much sharper when the ``product rule'' was applied. However, fig.2 of \cite{Ziff2009} shows that although $\Delta/N$ scales as a power of the system size $L$, as it was expected for the ``product rule'', the same is true for classical bond percolation! Of course, the scaling exponent is different since the explosive transition is indeed much sharper than the classical. But instead of the expected result $\Delta/N \rightarrow c$ where $c$ is a constant, one observes a clear power law decrease of $\Delta/N$ with increasing system size. This is a very important problem as it implies that according to the $\Delta$ criterion, classical bond percolation on a square lattice should also be a first-order transition which is absolutely not correct. Thus, the criterion produced a ``false'' positive in the case of lattice bond percolation. 

Initially, it was believed that this discrepancy was the result of finite size effects, i.e. that the observed scaling would disappear if larger systems were simulated and that the desired behavior would be recovered in the infinite system limit. But one should note that the systems simulated in  \cite{Ziff2009,Ziff2010a} are already rather large and simulation of much larger systems is prohibited due to CPU limitations.

The $\Delta$ criterion is not the only one used to determine the order of the transition. The vast majority of the published papers combine different scaling relations at the critical point, such as $P_{max}(L,p_c) \sim L^{-\beta/\nu}$  and $M'_{2}(L,p_c) \sim L^{\gamma/\nu}$, the scaling of the susceptibility $\chi_{max}(L,p_c)$, the cluster size distribution at $p_c$, hysteresis loops (presented in subsequent section) and the scaling of the average size of the largest jump of the order parameter \cite{Nagler2011, Manna2011,Chen2011a, Manna2012,Schrenk2012a,Chen2012,Cho2013,Chen2013b}. According to the last measure, the largest jump of the order parameter $\Delta C_{max}$, which scales as $N^{1-x}$, can be used to identify if a system exhibits a strong ($\Delta C_{max} \sim N$) or weak ($\Delta C_{max} \rightarrow 0$ as $N \rightarrow \infty$) discontinuous percolation transition.
However, the search for an easy to implement numerical criterion (similar to the $\Delta$ criterion) which will allow a safe distinction between the first and second order transitions in EP is still an important open problem.

The explosive character of the percolation transition under the product rule had already succeeded in capturing the interest of the scientific community.
Radicchi and Fortunato \cite{Radicchi2009,Radicchi2010a} have studied the effects of the ``product rule'' on Scale-free networks i.e. graphs where the degree distribution follows a power law.
More specifically, they study scale-free networks constructed via a cooperative Achlioptas growth process. Links between nodes are introduced in order to produce a scale-free graph with given exponent $\lambda$ for the degree
distribution, but the choice of each new link is done through the application of the ``product rule''. They find that the constructed networks are rather different from ``normal'' scale-free networks i.e. when links are introduced just randomly. They find that the ``product rule'' to a phase transition with a non-vanishing percolation threshold already for $\lambda > \lambda_c \sim 2.2$ while for random scale free networks $\lambda >3 $ is required.
More interestingly, they report that the transition is continuous when $\lambda< 3$ but becomes discontinuous when $\lambda>3$. 
They \cite{Radicchi2010a} also repeat the calculations of Ziff for explosive bond percolation on lattices using a slightly modified method of analysis. Again the false positive of the $\Delta$ criterion is observed and attributed to finite size effects - a rather plausible assumption based on what was known at the time this work was published.

At the same time, a similar work was done \cite{Cho2009}. The authors estimated the critical point $\lambda_c$ in the range (2.3,2.4). For $2< \lambda < \lambda_c$, $p_c(N \rightarrow \infty) \rightarrow 0$ which points to a continuous percolation transition. For $ \lambda_c < \lambda < 3$, $p_c$ remains finite in the thermodynamic limit and the transition was determined to be ``explosive''. They also found an effective $\lambda$ value at $p_c$ different from the theoretical one. In order to explain the differences, they proposed a mechanism which was based on the competition between the natural tendency of the hubs to participate more frequently in the formation of the giant component and the suppression effect of the Achlioptas Process. However, the authors in \cite{Cho2009} stress the differences between their model and the one used in \cite{Radicchi2009}.

Obviously, explosive percolation is a topic of vast theoretical interest. There are three very interesting scientific efforts, however, which use explosive percolation as a means to probe and understand scientific problems of more practical interest. In 2010, \emph{Kim et al.} \cite{Kim2010}, used the ``Achlioptas Process'' to investigate the percolation properties of a system of single-walled nanotube bundle with uniform diameter. Placing sticks of length $l$ on a square area of size $L (L>>l)$ and suppressing the formation of the largest cluster, they found that the growth mechanism of the real -world system is similar to the loopless bond percolation on $2d$ square lattices. Based on the hysteresis loop formed between the forward and inverse process (i.e. the removal of sticks in order to break the largest cluster), they concluded that the transition is of first-order.

In a 2011 article entitled ``Using explosive percolation in analysis of real-world networks'', Pan et al. \cite{Pan2011a}  apply a variant (not the ``product rule'' per se) of the explosive percolation procedure to large real-world networks. They show that the universality class of the  percolation transition depends on the structural properties of the network as expected , and also on  the number of unoccupied links considered for comparison in the procedure.
Most importantly they apply this process not only on model networks but also on real social networks and find that the percolation clusters close to the critical point are related to the community structure.

Again in 2011, another impressive article by Rosenfeld et al. \cite{Rozenfeld2010} appears. It is entitled ``Explosive percolation in the human protein homology network''. There the authors show that the emergence of a spanning cluster in the Human Protein Homology Network has features
similar  to an explosive transition and is markedly different from classical random percolation. To be more specific, the authors, based on the structural properties of this protein network in which homologous proteins form dense clusters (i.e. high modularity), propose a deterministic model in which the clusters are kept isolated by forcing them to double in size via merging in each time step. At $p=1$, an abrupt transition occurs leading to the formation of a giant connected component. The authors stress that the sharp transition observed is an indication that the real-world system exhibits Achlioptas Processlike features. 
Finding other ``real world'' processes that may be related to EP transitions is still a very important open problem, since up to now research on EP is driven more from a theoretical scientific curiosity than from application oriented or engineering needs.

\section{Related models and generalizations}

The appearance of the ``explosive percolation''  model \cite{Achlioptas2009a} initiated considerable efforts for the determination of the underlying mechanism responsible for the formation of the giant component. To answer this question, \emph{Friedman et al.} \cite{Friedman2009} proposed the notion of ``powder keg''. According to it, before the onset of the transition, a collection of small-sized clusters in a specified range is formed, composing the ``keg''. Their total mass remains constant as we approach the thermodynamic limit. Between them, bonds are drawn, leading to the formation of the giant component.  The authors concluded that the growth rules are only responsible for the formation of the ``powder keg'' and do not affect the order of the phase transition.

The existence of second - order phase transition features, like for example a power law cluster size distribution of the EP model, led many authors to search for new variants that exhibit a clear first-order transition. In order to achieve the control over the bonds, \emph{Moreira et al.}\cite{Moreira2010a} proposed a Hamiltonian of the form 
\begin{equation}
H(G)=\sum_{i \in \textbf{C}}{s_i^2 + l_is_i^{\beta}},
\label{hamiltEq}
\end{equation}
where $s_i$ is the number of vertices in cluster $i$ and $l_i$ the number of redundant bonds added to this cluster.
Depending on the values of the control parameter $\beta$, they observed that for small values ``redundant'' bonds dominate over ``merging'', leading to fully connected clusters with the transition being second-order. For large values of $\beta$, the opposite procedure takes place, a tree-like structure emerges and the system undergoes a first-order transition in the thermodynamic limit. In this framework, they provided a connection between explosive percolation and equilibrium statistical mechanics, considering the case where $\beta \rightarrow \infty$. The key features of their approach are that the homogenization of the cluster sizes and the domination of ``merging'' bonds over ``redundant'' ones, are sufficient properties for the system to exhibit ``explosive'' transition, the  results being independent of the dimensionality. 

In an attempt to explore the behavior of \emph{Achlioptas - like} models, \emph{Andrade et al.} \cite{Andrade2011} have investigated the structural and transport properties when a $best  - of -m $ rule is applied on a square lattice. According to this model, one selects the bond which minimizes the product of the cluster sizes to be merged, out of $m$ candidate ones. As $m$ increases, the ``explosive'' transition becomes more pronounced and at $m \rightarrow \infty$, we recover a clear first-order transition.

In  \cite{Araujo2011}, a ``hybrid'' model on a square lattice is considered, where the $best - of -m$  product rule \cite{Andrade2011} is selected with probability $q$ and the ordinary percolation model with probability $1-q$. For $q>q_t$ ($q_t = 0.51 \pm 0.01$), the transition falls in the ordinary percolation universality class while for $q<q_t$ the transitions are first-order. The authors have calculated the critical exponents at $q_t$ which are different from those of the ordinary percolation. They also concluded that, by increasing the value of $m$, the transition becomes more ``explosive'' (being totally ``discontinuous'' at $m \rightarrow \infty$ \cite{Andrade2011}) and the value of $q_t$ shifts to larger values.

A different approach is followed in  \cite{Araujo2010,Schrenk2011a}. By keeping the clusters narrowly distributed and suppressing the formation of the giant component, the authors manage to produce a first-order phase transition. They present numerical results for the square lattice in two \cite{Araujo2010}, three and higher dimensions \cite{Schrenk2011a}, based on a relation of the form 
\begin{equation}
min\left\{1,\exp{\left[-\alpha \left(\frac{s-\bar{s}}{\bar{s}}\right)^2 \right]}\right\},
\label{Araujo1}
\end{equation}
where $s$ is the size of the cluster and $\bar{s}$ the average cluster size, after the insertion of the chosen bond (from a pool of empty ones) and $\alpha$ the control parameter, with  $\alpha \leq 0$  recovering the classical percolation universality class and  $\alpha > 0$ exhibiting a first-order transition. 

For $d \geq 2$ and $\alpha>0$, the model (eq. \ref{Araujo1}) leads to a discontinuous transition, which is verified by the behavior of various quantities in the thermodynamic limit at $p_c$  (largest jump $J$, maximum of the second moment of cluster size distribution $M_2'$, bimodal cluster-size distribution, standard deviation of the order parameter $\chi_{\infty}$, see \cite{Schrenk2011a}). The clusters at the transition point have fractal perimeter $d_A$ which converges to $d$ with increasing dimension. Finally, the authors propose a lower bound for the percolation threshold to be $p_c = 1/d$ , which reaches the value of $1/N$ at infinity  ($N$ the number of sites).

In \cite{Chen2011a}, the percolation properties of the Bohman - Frieze - Wormald (BFW)  model \cite{BFW2004} are investigated on networks. The authors find, both for restricted (sampling edges that merge only different clusters) and unrestricted case, that near the critical point, there exist multiple ``macroscopic'' clusters. Depending on the control parameter $\alpha$, they arrive to different conclusions: (a) for the case of unristricted sampling and $\alpha = 1/2$, only $2$ macroscopic clusters appear, which remain stable (do not merge) till the end of the process and (b) for the restricted case and all the values of $\alpha$, several multiple ``giant'' clusters (their number depending on $\alpha$) coexist, which collapse to a single ``giant component'' at the critical point (which increases with decreasing $\alpha$). The transition, in all cases of (b), is discontinuous (being more pronounced as $\alpha$ decreases).

In \cite{Schrenk2012a}, the percolation properties of the Bohman - Frieze - Wormald (BFW)  model are investigated on $2D$ (square and triangular) and  $3D$ (simple cubic) lattices. Using the same quantities as in \cite{Schrenk2011a},  the authors find that these quantities are independent of lattice size, pointing to a discontinuous transition. The clusters are homogeneous, compact and have fractal perimeter. In $3D$, $P_{\infty}$ exhibits a stepwise evolution, indicating the existence of multiple stable clusters . This result is attributed to the existence of redundant bonds. 

They have also tested the case of the \emph{restricted BFW model} where only merging bonds are present and the evolution of clusters is tree-like. 
For all considered dimensions ($d=2-7$) , the transition is discontinuous, with no ``plateaus'' (as in the unrestricted case), and $p_c \simeq \frac{1}{d}$ (confirming \cite{Schrenk2011a}). The authors conclude that the homogenization of the clusters and the predominance of merging bonds over redundant ones are sufficient to induce discontinuity in a system (see also \cite{Araujo2010,Schrenk2011a}).

In an effort to propose a unifying scheme for the ``explosive'' percolation models, \emph{Cho et al.} introduced a Spanning Cluster Avoiding (SCA) model \cite{Cho2013}, based on the differentiation between bridge and non-bridge bonds and keeping the former suppressed. The authors have identified a tricritical point $m_c$ which depends on the dimension $d$ of the system (for $d<d_c=6$). For $m<m_c$, the percolation transition ($PT$) is continuous for $N \rightarrow \infty$ and the critical point $t_{cm}(N \rightarrow \infty) \rightarrow t_c$ (the ordinary percolation critical point). For  $m>m_c$, $PT$ is discontinuous for $N \rightarrow \infty$ and the critical point $t_{cm}(N \rightarrow \infty) \rightarrow 1$. At $m=m_c$, $t_c< t_{cm_c}(N \rightarrow \infty) <1$. 

The case is more clear for $d \geq d_c$. For fixed $m$, $t_{cm}(N \rightarrow \infty) \rightarrow t_c$, thus $PT$ is continuous. However, if $m \sim lnN $, $t_{cm}$ is finite ($ \neq t_c, 1$) and $PT$ is discontinuous. Also, if $m$ increases faster than $lnN$, they recover the case of bridge percolation \cite{Schrenk2012b}. In all cases, homogeneous and compact clusters with fractal perimeters are formed for the ``discontinuous'' regimes (i.e. $m>m_c$ for $d<d_c$). The authors have also examined the product rule case (PR) and concluded that the PT is continuous for  $N \rightarrow \infty $ and fixed value of $m$, for all $d$.  However, if $m$ increases faster than $lnN$, the PT is discontinuous.

Finally, the authors present a criterion for distinguishing the order of a PT. Specifically, by taking the scaling of the maximum rate of change of the order parameter $dG_m(t)/dt$ ($t$ is the number of links present in the system) with the system size $N$, and applying the relation $(1-x)d_{BB}/d$   ($d_{BB}$ is the fractal dimension of bridge bonds), they argue that for $x<1$  the transition is discontinuous, whereas for $x=1$ the transition is continuous.

The above results, compared to those presented in the next section, indicate that there is an open debate about the order of the phase transition.


\section{Theoretical Results}


Explosive percolation was treated initially as being a discontinuous phase transition. A milestone in the question whether explosive percolation is continuous or not came in 2010, when da Costa et al. \cite{daCosta2010} presented a mathematical approach, based on both analytical arguments and numerically solved rate equations. The combination of these two contradicted the interpretation of simulation results presented until then by others.
They used a representative model, slightly altered than that of \cite{Achlioptas2009a}, yet keeping the same key elements. In their model they consider the clusters two randomly chosen nodes belong to, and compare them. They select only the node belonging to the smallest cluster and repeat the same process in order to select a second node (through a new set of nodes belonging to a new set of clusters). Then, these two are linked, and the two clusters the two nodes belong to are merged. This model is similar to the original product rule found in \cite{Achlioptas2009a} in the sense that, in merging, the minimal clusters are selected from two possibilities.

In fact, to further point out the existence of finite size effects, they showed that even for a system of $2\times10^9$ nodes, it is not possible by simulations only to validate or rule out discontinuity. Specifically, when plotting $S(t)$ (time was used to describe the total number of added links in the system) a discontinuity seems to appear appear at the critical point $t_c$. Upon a closer inspection a power law $S\propto(t-t_c)^\beta$ is evident, pointing to a continuous transition.

A numerical solution gives a very small value of $\beta$, $\beta=0.0555(1)$, close to $1/18$, for $t_c=0.9232$. It also shows that the evolution of the size distribution over time, $P(s,t)$, in an explosive percolation model presents peaks below $t_c$ and depend on system size. On the other hand, ordinary classical percolation yields peaks symmetrical with respect to $t_c$, dependless of the system size (for large enough sizes).

The analytical treatment confirms these results and helps calculate the fractal and upper critical dimensions, thus determining the finite size effect. The upper critical and fractal dimensions found were much smaller than those of ordinary percolation. The relative simplicity of the model also allowed for a $best - of -m $ rule similar analysis, which followed shortly after in other papers \cite{Andrade2011, Araujo2011}.

Overall, this paper suggested for the first time that this transition is actually continuous, yet with very small critical exponent values. In order though to achieve such a result, it was assumed in the analytical treatment that the transition is indeed continuous, and was later on verified in the same paper. Therefore, even authors that supported the existence of a continuous phase transition \cite{Riordan2011a}, viewed this manuscript as a heuristic and not as a proof.

Two more papers \cite{Grassberger2011a,Riordan2011a}, that followed shortly after and were published almost simultaneously, supported this claim. 

In \cite{Grassberger2011a} four Achlioptas-type processes were studied: $i)$ the original product rule \cite{Achlioptas2009a}, $ii)$ the same rule on 2D lattices, $iii)$ the adjacent rule \cite{DSouza2010b}, and $iv)$ the model of \cite{daCosta2010}. By finding the distribution of the order parameter and the critical exponents in all cases, they show that all models have an unusual finite size behavior with a nonanalytic scaling function. They also show that $\beta$ is indeed very small and thus it is possible to be confused in finite size systems. The values obtained support their claim that each of the four processes belongs in a different universality class, yet none shows a discontinuous transition.

Finally, Riordan and Warnke \cite{Riordan2011a} have shown that all ER model based Achlioptas processes have continuous phase transitions. Using their approach, one can use any rule based on picking a fixed number of vertices to obtain a continuous transition. On the other hand, they have made clear that other related models in which the number of nodes sampled may grow with the network size, can exhibit a discontinuous phase transition. As an example, they use the Smallest Distinct Components (SDC) rule, which is similar in practice with the $best  - of -m $ rule used along with the method of \cite{daCosta2010}. It simply takes into account $l$ random vertices out of a total $n$, and join those two vertices that belong to the SDC. The difference is that $l\to\infty$ and it also depends on $n$, Thus, if the total number of vertices $n\to\infty$, it is possible to obtain a discontinuous transition.

Due to the numerical difficulties associated with the investigation of EP processes, the identification of simple models where exact results are possible is of high importance and an open problem for future research.
 
\section{Other Efforts}

\begin{figure}
\begin{center}
\includegraphics[angle=0,width=12cm]{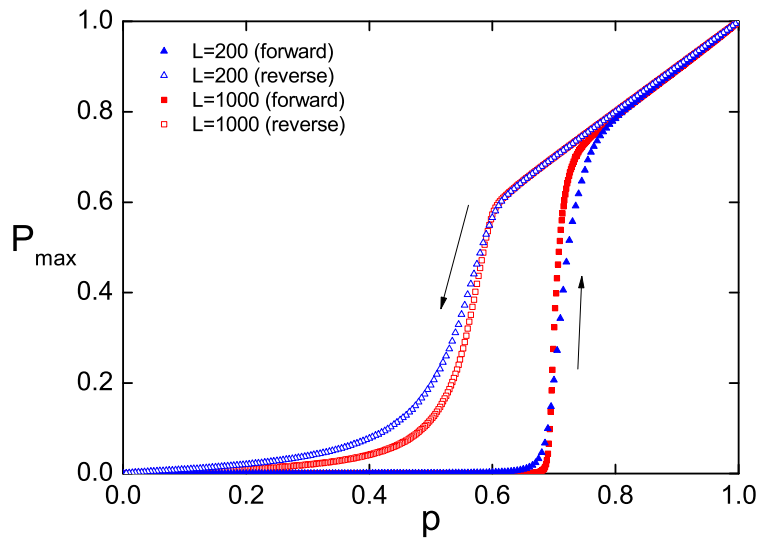}
\end{center}
\caption{(Color online) Plot of fraction of the largest cluster size, $P_{max}=S_{max}/N$, as a function of $p$ for the ``Sum rule'' for explosive percolation. Full symbols indicate the forward and empty symbols  the reverse procedure (see also the arrows). Triangles  are for $200\times200$ and squares for $1000\times1000$ lattices. The hysteresis loop for the different curves of the same system encloses an area which is proven to decay as $L$ is increased, implying that it vanishes in the thermodynamic limit and, thus, pointing to a continuous phase transition.}
\label{Schema2}
\end{figure}

A significant part of the research on explosive percolation has focused in an effort to determine the order of the phase transition. However, the idea of introducing a selection rule which is more complex than the classical case has several additional important effects. For example, in classical percolation there is actually no difference between a direct process (i.e. adding a bond randomly to an empty ``lattice'') and the reverse process (randomly removing bonds from an initially fully occupied lattice). By applying a seemingly simple rule (such as the product rule for example) there is a meaningful distinction between a forward and a reverse process. In \cite{bastas2011explosive}, Bastas et al. have introduced this concept of forward and reverse EP processes and investigated for the presence of hysteresis loops (fig. \ref{Schema2}). They have demonstrated that while the reverse process is different from the direct process for finite size systems, the two cases become equivalent in the thermodynamic limit. This was based on the scaling of the area enclosed between the lines of the $2$ processes (see fig. \ref{Schema2}, which was proven to decay as $L$ increased. This findings pointed to the conclusion that the area tends to $0$ in the thermodynamic limit. Since hysteresis is typically associated with first-order transitions, this work indirectly shows that product rule EP processes exhibit only finite size hysteresis and behave like continuous transitions in the thermodynamic limit.

In \cite{Choi2011} site percolation under Achlioptas process (AP) in a 2D lattice has been studied. From the scaling of the order parameter $P_{\infty}(p)$ (Eq.\ref{eq1}) with the system size $s$, Choi et al. found that $P_{\infty}(p)$ has a stable hump when $p$ is lower than $p_c$, which indicates that below $p_c$ there are a lot of microscopic but stable clusters. As $p$ approaches $p_c$ , $P_{\infty}(p)$ has a very robust power-law regime followed by the hump. They also, obtained a value for the exponent $\delta$, $\delta = 0.9$, $[P_{\infty}(p_c) = s^{-\delta}]$ and since $\delta$ is below unity there should be a cutoff in the possible cluster size for $p$ unlike that of classical percolation. Thus, to generate a macroscopic cluster there should be a discontinuous jump in the limit $L \rightarrow \infty$ and the transition becomes discontinuous. Of course this result is at odds with the exact solutions presented at \cite{Riordan2011a} and should be reconsidered. Hysteresis measurements for this transition were once more investigated in the same paper. They resulted in the fact that in 2D lattices there is a gap between the critical density of sites ($p_c$) in the two different ways of constructing the system (forward and reverse). In contrast to \cite{bastas2011explosive} they claim that in the thermodynamic limit this gap seems to go to a constant value instead of going to zero (continuous phase transition), resulting in an indication that site percolation in 2D lattices under ``Achlioptas procceses'' is a discontinuous phase transition. This is again at odds with both \cite{bastas2011explosive} and \cite{Riordan2011a}.

On the other hand, Liang et al.\cite{Tian2012} implemented the Achlioptas process in random graphs, scale-free networks, and in 2D lattices. By examining the order parameter distribution histogram at the percolation threshold, they found that two well-defined Gaussian-like peaks coexist, which represent the non-percolative phase and percolative phase, respectively. The two peaks gradually get close to each other with an increasing system size, and in the thermodynamic limit they merge at the transition point of the order parameter. These observations suggest the explosive percolation is a continuous phase transition with first-order-like finite-size effect.

Another version of Achlioptas process called ``Generalized Achlioptas process'' was used in \cite{Fan2012} on Random networks. In this procedure, Jingfang et al. chose two independent possible bonds out of which they retain with a probability $p$ the one that minimizes the product of the connecting clusters sizes that are merged by this bond. If $p=\frac{1}{2}$ then this model gives the classical percolation. Furthermore, for $p=1$ the model is equivalent to Achlioptas process (AP). By implementing this generalization Jingfang et al. found that in the entire range of p [0.5,1], the percolation transition is a continuous one. For $0.5<p<0.8$ the critical exponents of the order parameter are unchanged and the phase transition remains in the same universality class. A different behavior arises for $p>0.9$, where the exponents at critical point vary with $p$ and the universality class of phase transitions depends on p.
On a similar context, Liu et al.\cite{Liu2012} studied the AP for a 2D square lattice and they resulted that the universality class of this continuous phase transition in such systems is always dependent of the $p$ parameter.

\begin{figure}
\begin{center}
\includegraphics[angle=0,width=12cm]{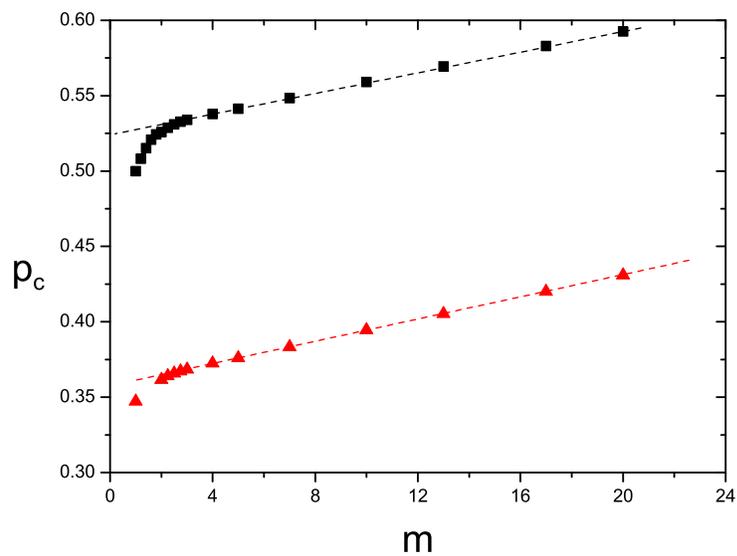}
\end{center}
\caption{(Color online) The delay of the critical density ($p_c$) of bonds in 2D square (black squares) and triangular (red triangles) lattices as a function of the number of candidate bonds $m$.   }
\label{fig2}
\end{figure}

Another generalization of the "Achlioptas process" was studied in \cite{Paris2013}. This generalized procedure of product rule was implemented at the bond percolation problem in 2D and 3D lattices. In this method, Giazitzidis at al., probed a number of candidate bonds ($m$) and retained only the one that minimizes the product of the connecting cluster sizes as done in previous publications\cite{Andrade2011}. Their results showed, as expected, that as the number of the candidate bonds grows, the expected value of $p_c$ rises. For $m=1$ the classical percolation transition occurs, while the "Achlioptas process" is for $m=2$. This method was implemented for $2<m<20$ and a delay of the criticality depending on the $m$ parameter was found. They also find out that this delay is proportional to the $m$ parameter and more specifically, the same system dimensionality results to equal slope of this critical density increase. Figure \ref{fig2} shows this proportional increase of the critical density of bonds, as a function of the number of candidate bonds been chosen. The same study explored another variant of the "explosive percolation" model. In this model one starts with an empty lattice and bonds to be added are picked randomly. Candidate bonds are retained with probability $p=\frac{1}{S}$, where $S$ is the size of the cluster that this bond merges. Adding bonds in a 2D lattice with a probability inversely proportional on the cluster size, leads to an "explosive" behavior of the system. Nevertheless, as the variance of the largest cluster diverges depending on the density of bonds $p$, this phase transition is clearly a continuous one. The authors also compare these two models and extract useful results.  

In the context of presenting a full review of the work done in the topic of explosive percolation, we present below two detailed summary tables. Table \ref{table1} gathers a large part of the percolation threshold and critical exponent values found in the literature, while table \ref{table2} lists most of the methods categorized by system properties.

\newgeometry{top=1.8cm,bottom=1.8cm,left=1.5cm,right=1.5cm}
\begin{landscape}
\begin{flushright}
\scriptsize
\begin{tabular}{ |m{1.5cm} | m{4.7cm} | m{4.7cm} | m{4cm} | m{2.3cm} | m{1.3cm} | m{1.3cm} |m{0.8cm}|} \hline
Topology & Method & $p_c$ & \multicolumn{4}{  c | }{Critical Exponents} & ref \\ \cline{4-7}
& &  & $\beta/\nu$ & $\gamma/\nu$  & $\nu$ & $\tau$  & \\ \cline{1-8}
Networks & Achlioptas product rule & $0.888...$ & & & & & \cite{Achlioptas2009a} \\ \hline
Regular lattices & Achlioptas product rule (bond percolation) & $t_c/N \simeq 0.9925$ & & & & & \cite{Ziff2009} \\ \hline
Networks & Achlioptas product rule (bond percolation) on SFNs with given $\gamma$ & fig.3  & fig.4 & fig.4 & & & \cite{Radicchi2009} \\ \hline
All & Achlioptas model & \multicolumn{5} {c|} {TABLE I} & \cite{Radicchi2010a}  \\ \hline
Networks & Local models: ``Adjacent edge'' (AE) and ``Triangle rule'' (TR) & $t_c=0.796$ (AE)  $t_c=0.848$ (TR) (using the procedure proposed in \cite{Achlioptas2009a})  & & & & $2.1$ &  \cite{DSouza2010b}\\ \hline
Networks and lattices & link occupation probability $\sim (s_is_j)^{\alpha}$ & fig3b for square lattices, fig5b for random networks (GM) & & & & & \cite{Manna2011} \\ \hline
2D lattices & Largest Cluster (LCM) and Gaussian (GM) models & $0.632(20)$ (LCM), $0.56244(6)$ (GM) & & & & & \cite{Araujo2010} \\ \hline
Square lattice & Achlioptas product rule (bond percolation) & $0.526565(5)$ & $0.06(1)$ & $1.90(1)  (M_2/N)$ or $2 (M'_2/N)$ & $1.04(1)$ or $1 (\xi)$ & $2.025(10)$ & \cite{Ziff2010a} \\ \hline
Networks & Achlioptas-like model & \multicolumn{5} {c|} {TABLE I} & \cite{daCosta2010}  \\ \hline
2D lattices & MC - m rule & inset of fig.4 & & & & & \cite{Andrade2011} \\ \hline
Real world networks & Best of $m$ sum rule & & MPC: $\sim 0.42$ (random), $\sim 0.14$ (MC-2), $\sim 0.03$ (MC-10) \par CA: $\sim 1.15$ (random), $\sim 0.94$ (MC-2), $\sim 0.13$ (MC-10) \par SCA: $\sim 0.7$ (random), $\sim 0.31$ (MC-2), $\sim 0.06$ (MC-10) & & & & \cite{Pan2011a} \\ \hline
Networks and lattices & $4$ models for explosive percolation : ``product rule'' (PR), PR on $2D$ lattices (2d), ``adjacent edge'' (AE), ``da Costa'' (CDMG) & Table I, first row & & & & &  \cite{Grassberger2011a}\\ \hline
2D lattices & Achlioptas product rule  & $0.768(3)$ & $0.011(2)$ & $1.98(1)$ & & $0.9(2)$ &  \cite{Choi2011}\\ \hline
Lattices ($d=3$ to $\infty$) & Gaussian model & Table I & & & & & \cite{Schrenk2011a} \\ \hline
2D lattices & Achlioptas model & $0.695$ (``Sum rule'' - SR), $0.756(6)$ (``Product rule''-PR) & $0.001$ (SR), $0.04(2)$ (PR) & & & &  \cite{bastas2011explosive}\\ \hline
Lattices & BFW model & $1.000(2)$ 2D , $0.500(1)$ 2D, $0.333(2)$ 3D, $0.248(2)$ 4D, $0.198(3)$ 5D, $0.165(4)$ 6D,$0.141(9)$ 7D (tree-version) & & & & & \cite{Schrenk2012a} \\ \hline
2D square lattices & Generalized Achlioptas model & Table $1$ & Table $1$ & & Table $1$ ($1/\nu$) & &  \cite{Liu2012}\\ \hline
Bethe lattice & Achlioptas process & & $0.05(5)$ & $1.00(1)$ & &  &  \cite{Chae2012}\\ \hline
Networks & Generalized Achlioptas model & Table I & Table I & & Table I ($1/\nu$) & &  \cite{Fan2012}\\ \hline
Networks & Achlioptas site percolation (best - of -m rule) & $t_c(k,m) \sim k^{-\lambda(m)}$, $T_c(k) \sim k^{-\tau}$, k=average degree, $-1 \leq \tau \leq -0.5$  & $\beta(m) \sim m^{-1.1}$ (the hyperscaling relation holds) & & &  &  \cite{Qian2012}\\ \hline
Networks & Achlioptas process on directed networks & Table I & Table I & & Table I ($1/\nu$) &  &  \cite{Squires2013}\\ \hline
Networks & Achlioptas process on growing networks & $0.5149(1)$ & $0.20(1)$ &  & $0.40(1)$ ($1/\nu$) & $2.24(1)$  &  \cite{Yi2013}\\ \hline
Networks & degree-dependent linking probability models & $<k>_c \rightarrow 2$ as $\alpha \rightarrow \infty$ & 1 & & & &\cite{Hooyberghs2013b} \\ \hline
\end{tabular}
\captionof{table}{Collection of the values of $p_c$ and the critical exponents, based on the chronological order of appearance in the literature.}
\label{table1}
\end{flushright}
\end{landscape}
\restoregeometry

\begin{flushleft}
\scriptsize
\begin{tabular}{ |m{1.5cm} |c |m{10cm}|} \hline
\multirow{4}{*}{Lattices} & 1D &  1D with small world Bonds \cite{Boettcher2012} \\ \cline{2-3}
 
& 2D & Explosive Ising \cite{Angst2012}, k-core tricriticality \cite{Cao2012}, Bond-Site \cite{Choi2012}, ``product rule'' \cite{Radicchi2010a}, ``Explosive'' percolation \cite{Andrade2011}, ``Largest cluster'' (LC) and ``Gaussian model'' \cite{Araujo2010}, hybrid model (selection between random and ``best-of-$10$'' (Achlioptas) rule) \cite{Araujo2011}, ``Sum'' and ``Product'' Achlioptas site percolation rule \cite{bastas2011explosive}, ``Product'' site percolation \cite{Choi2011}, Generalized Achlioptas model \cite{Liu2012}, q-state Potts model \cite{Lv2012}, biased link occupation rule (Achlioptas-like process) \cite{Manna2011}, product rule \cite{Ziff2009,Ziff2010a}, Bohman-Frieze-Wormald (BFW) model \cite{Schrenk2012a}, explosive percolation \cite{Tian2012}\\ \cline{2-3}
 
& 3D & ``product rule'' \cite{Radicchi2010a}, Bohman-Frieze-Wormald (BFW) model\cite{Schrenk2012a}, Gaussian model \cite{Schrenk2011a} \\ \hline

\multirow{4}{*}{Networks} & Random Graphs & k-core tricriticality  \cite{Cao2012}, ``product rule'' \cite{Radicchi2010a}, random - explosive partial product rule \cite{Chi2012}, ER - cluster aggregation model  \cite{Cho2010a}, random percolation and ``product'' rule \cite{Cho2010a}, ``product rule'' - ``sum rule'' - ``suppression principle'' \cite{Cho2011}, Original Achlioptas percolation model \cite{Achlioptas2009a}, da Costa explosive percolation model \cite{daCosta2010}, Adjacent edge (AE) and Triangle rule (TR) local cluster model \cite{DSouza2010b}, Generalized Achlioptas model \cite{Fan2012}, Explosive percolation \cite{Lee2011}, biased link occupation rule (Achlioptas-like process) \cite{Manna2011}, Achlioptas process on growing networks \cite{Yi2013}, Hamiltonian model \cite{Moreira2010a}, dense percolation \cite{Veremyev2012}, explosive percolation \cite{Tian2012}, Discontinuous ER-like models \cite{Panagiotou2011,Panagiotou2013} , Explosive site percolation \cite{Qian2012}\\ \cline{2-3}
 
& Scale free networks & ``Achlioptas process'' (AP) rule \cite{Cho2009} , ``product rule'' \cite{Radicchi2010a}, explosive synchronization transition \cite{Gomez2011}, explosive percolation \cite{Tian2012}, explosive percolation \cite{Radicchi2009}\\ \cline{2-3}
 
& Others &  Bohman-Frieze-Wormald (BFW) model \cite{Chen2011a}, BFW model \cite{Chen2012},  BFW model (supercritical regime) \cite{Chen2013b}, ``Devil's Staircase'', ``Nagler-Gutch'', modified ER and BFW models (supercritical regime) \cite{Chen2013arxiv},ER - cluster aggregation model  \cite{Cho2010a}, spectral analysis of Smallest Cluster (SC)-Gaussian model \cite{Chung2013}, hierarchical networks \cite{Kachhvah2012}, \\ \hline

Other & Other & Bethe lattices \cite{Chae2012}, diffusion - limited cluster aggregation (DLCA) model \cite{Cho2011b}, nanotube-based systems\cite{Kim2010}, Gaussian model in $d > 3$ \cite{Schrenk2011a}, best-of-m rule on real-world networks \cite{Pan2011a}, Weakly explosive percolation \cite{Squires2013}, deterministic explosive percolation process \cite{Rozenfeld2010} , degree-correlated network growth models with fixed $N$ \cite{Hooyberghs2013b} \\ \hline
\end{tabular}
\captionof{table}{List of the methods related to ``explosive percolation'' or other models which exhibit ``discontinuous'' transition from the existing literature.}
\label{table2}
\end{flushleft}

\section{Conclusions}

Explosive percolation is a very interesting and promising scientific problem within the larger field of percolation. It holds considerable potential to advance our knowledge of phase transitions and critical phenomena.
The fact that a slight modification of the classical selection rules can significantly alter the universality class of the underlying phase transition is impressive. Moreover, it seems that in extreme cases the order of transition itself may change as well.
Studying explosive percolation processes has revealed a variety of unexplored and poorly understood geometric phase transitions with unusual finite size scaling.  
Their study is expected to lead to better computational tools and methods of understanding scaling, as well as distinguishing between the nature of phase transitions. It is also expected to lead to a deeper understanding of critical phenomena by allowing us to study relatively simple models which, nevertheless, exhibit very rich and unexpected physical properties.


\section*{Acknowledgements}
N. B. acknowledges financial support from Public Benefit Foundation Alexander S. Onassis.







\bibliographystyle{elsarticle-num}
\bibliography{ExplosivePercolation}

\begin{thebibliography}{10}
\expandafter\ifx\csname url\endcsname\relax
  \def\url#1{\texttt{#1}}\fi
\expandafter\ifx\csname urlprefix\endcsname\relax\def\urlprefix{URL }\fi
\expandafter\ifx\csname href\endcsname\relax
  \def\href#1#2{#2} \def\path#1{#1}\fi

\bibitem{Bunde_book}
A.~Bunde, S.~Havlin, Fractals and Disordered Systems, Springer-Verlag, Berlin-
  Heidelberg, 1996.

\bibitem{Stauffer_Book}
D.~Stauffer, A.~Aharony, Introduction to Percolation Theory, Taylor \& Francis,
  London, 1994.

\bibitem{Schrenk2013b}
K.~J. Schrenk, N.~Pos\'e, J.~J. Kranz, L.~V.~M. van Kessenich, N.~A.~M.
  Ara\'ujo, H.~J. Herrmann,
  \href{http://link.aps.org/doi/10.1103/PhysRevE.88.052102}{Percolation with
  long-range correlated disorder}, Phys. Rev. E 88 (2013) 052102.
\newblock \href {http://dx.doi.org/10.1103/PhysRevE.88.052102}
  {\path{doi:10.1103/PhysRevE.88.052102}}.
\newline\urlprefix\url{http://link.aps.org/doi/10.1103/PhysRevE.88.052102}

\bibitem{Achlioptas2009a}
D.~Achlioptas, R.~M. D'Souza, J.~Spencer,
  \href{http://dx.doi.org/10.1126/science.1167782}{Explosive percolation in
  random networks}, Science 323~(5920) (2009) 1453--1455.
\newblock \href {http://dx.doi.org/10.1126/science.1167782}
  {\path{doi:10.1126/science.1167782}}.
\newline\urlprefix\url{http://dx.doi.org/10.1126/science.1167782}

\bibitem{Adler1991}
J.~Adler, Bootstrap percolation, Physica A: Statistical Mechanics and its
  Applications 171 (1991) 453.

\bibitem{Parshani2011}
R.~Parshani, S.~V. Buldyrev, S.~Havlin, Critical effect of dependency groups on
  the function of networks, PNAS 108~(3) (2011) 1007.

\bibitem{Buldyrev2010}
S.~Buldyrev, R.~Parshani, G.~Paul, H.~Stanley, S.~Havlin, Catastrophic cascade
  of failures in interdependent networks, Nature Letters 464 (2010) 08932.

\bibitem{Parshani2010}
R.~Parshani, S.~Buldyrev, S.~Havlin, Interdependent networks: Reducing the
  coupling strength leads to a change from a first to a second order
  percolation transition, Phys. Rev. Lett. 105 (2010) 048701.

\bibitem{Gao2011}
J.~Gao, S.~Buldyrev, H.~Stanley, S.~Havlin, Networks formed from interdependent
  networks, Nature Physics 8 (2011) 2180.

\bibitem{Erdos1960}
P.~Erd\"{o}s, A.~R\'{e}nyi, On the evolution of random graphs, Publ. Math.
  Inst. Hungar. Acad. Sci. 5 (1960) 17.

\bibitem{Riordan2011a}
O.~Riordan, L.~Warnke,
  \href{http://dx.doi.org/10.1126/science.1206241}{Explosive percolation is
  continuous}, Science 333~(6040) (2011) 322--324.
\newblock \href {http://dx.doi.org/10.1126/science.1206241}
  {\path{doi:10.1126/science.1206241}}.
\newline\urlprefix\url{http://dx.doi.org/10.1126/science.1206241}

\bibitem{Ziff2009}
R.~M. Ziff, Explosive growth in biased dynamic percolation on two-dimensional
  regular lattice networks, Phys. Rev. Lett. 103 (2009) 045701.

\bibitem{Ziff2010a}
R.~M. Ziff, Scaling behavior of explosive percolation on the square lattice,
  Phys. Rev. E 82~(5 Pt 1) (2010) 051105.

\bibitem{Nagler2011}
J.~Nagler, A.~Levina, M.~Timme, Impact of single links in competitive
  percolation, Nature Physics 7 (2011) 265--270.

\bibitem{Manna2011}
S.~Manna, A.~Chatterjee,
  \href{http://www.sciencedirect.com/science/article/pii/S0378437110008460}{A
  new route to explosive percolation}, Physica A: Statistical Mechanics and its
  Applications 390~(2) (2011) 177--182.
\newline\urlprefix\url{http://www.sciencedirect.com/science/article/pii/S0378437110008460}

\bibitem{Chen2011a}
W.~Chen, R.~M. D'Souza, Explosive percolation with multiple giant components,
  Phys. Rev. Lett. 106~(11) (2011) 115701.

\bibitem{Manna2012}
S.~S. Manna, About the fastest growth of the order parameter in models of
  percolation, Physica A: Statistical Mechanics and its Applications 391 (2012)
  2833.

\bibitem{Schrenk2012a}
K.~J. Schrenk, A.~Felder, S.~Deflorin, N.~A.~M. Ara\'{u}jo, R.~M. {D'Souza},
  H.~J. Herrmann, {Bohman}-{Frieze}-{Wormald} model on the lattice, yielding a
  discontinuous percolation transition, Phys. Rev. E 85 (2012) 031103.

\bibitem{Chen2012}
W.~Chen, Z.~Zheng, R.~M. {D'Souza}, Deriving an underlying mechanism for
  discontinuous percolation, Eur. Phys. Lett. 100 (2012) 66006.

\bibitem{Cho2013}
Y.~S. Cho, S.~Hwang, H.~J. Herrmann, B.~Kahng,
  \href{http://dx.doi.org/10.1126/science.1230813}{Avoiding a spanning cluster
  in percolation models}, Science 339~(6124) (2013) 1185--1187.
\newblock \href {http://dx.doi.org/10.1126/science.1230813}
  {\path{doi:10.1126/science.1230813}}.
\newline\urlprefix\url{http://dx.doi.org/10.1126/science.1230813}

\bibitem{Chen2013b}
W.~Chen, J.~Nagler, X.~Cheng, X.~Jin, Z.~Shen, R.~M. {D'Souza}, Phase
  transitions in supercritical explosive percolation, Phys. Rev. E 87 (2013)
  052130.

\bibitem{Radicchi2009}
F.~Radicchi, S.~Fortunato, Explosive percolation in scale-free networks, Phys.
  Rev. Lett. 103~(16) (2009) 168701.

\bibitem{Radicchi2010a}
F.~Radicchi, S.~Fortunato, Explosive percolation: a numerical analysis, Phys.
  Rev. E 81~(3 Pt 2) (2010) 036110.

\bibitem{Cho2009}
Y.~S. Cho, J.~S. Kim, J.~Park, B.~Kahng, D.~Kim, Percolation transitions in
  scale-free networks under the achlioptas process, Phys. Rev. Lett. 103 (2009)
  135702.

\bibitem{Kim2010}
Y.~Kim, Y.~K. Yun, S.~H. Yook, Explosive percolation in a nanotube-based
  system., Phys. Rev. E 82~(6 Pt 1) (2010) 061105.

\bibitem{Pan2011a}
R.~K. Pan, M.~Kivela, J.~Saramaki, K.~Kaski, J.~Kertesz, Using explosive
  percolation in analysis of real-world networks, Phys. Rev. E 83~(4 Pt 2)
  (2011) 046112.

\bibitem{Rozenfeld2010}
H.~D. Rozenfeld, L.~K. Gallos, H.~A. Makse,
  \href{http://link.springer.com/article/10.1140/epjb/e2010-00156-8}{Explosive
  percolation in the human protein homology network}, The European Physical
  Journal B 75~(3) (2010) 305--310.
\newline\urlprefix\url{http://link.springer.com/article/10.1140/epjb/e2010-00156-8}

\bibitem{Friedman2009}
E.~J. Friedman, A.~S. Landsberg,
  \href{http://link.aps.org/doi/10.1103/PhysRevLett.103.255701}{Construction
  and analysis of random networks with explosive percolation}, Phys. Rev. Lett.
  103~(25) (2009) 255701.
\newline\urlprefix\url{http://link.aps.org/doi/10.1103/PhysRevLett.103.255701}

\bibitem{Moreira2010a}
A.~A. Moreira, E.~A. Oliveira, S.~D.~S. Reis, H.~J. Herrmann, J.~Andrade, Jr,
  Hamiltonian approach for explosive percolation, Phys. Rev. E 81~(4 Pt 1)
  (2010) 040101.

\bibitem{Andrade2011}
J.~S. Andrade, Jr, H.~J. Herrmann, A.~A. Moreira, C.~L.~N. Oliveira, Transport
  on exploding percolation clusters, Phys. Rev. E 83~(3 Pt 1) (2011) 031133.

\bibitem{Araujo2011}
N.~A. Ara\'{u}jo, J.~S. Andrade~Jr, R.~M. Ziff, H.~J. Herrmann,
  \href{http://link.aps.org/doi/10.1103/PhysRevLett.106.095703}{Tricritical
  point in explosive percolation}, Phys. Rev. Lett. 106~(9) (2011) 095703.
\newline\urlprefix\url{http://link.aps.org/doi/10.1103/PhysRevLett.106.095703}

\bibitem{Araujo2010}
N.~Ara\'{u}jo, H.~J. Herrmann,
  \href{http://link.aps.org/doi/10.1103/PhysRevLett.105.035701}{Explosive
  percolation via control of the largest cluster}, Phys. Rev. Lett. 105~(3)
  (2010) 035701.
\newline\urlprefix\url{http://link.aps.org/doi/10.1103/PhysRevLett.105.035701}

\bibitem{Schrenk2011a}
K.~J. Schrenk, N.~A.~M. Ara\'{u}jo, H.~J. Herrmann, Gaussian model of explosive
  percolation in three and higher dimensions, Phys. Rev. E 84~(4 Pt 1) (2011)
  041136.

\bibitem{BFW2004}
T.~Bohman, A.~Frieze, N.~Wormald, Avoidance of a giant component in half the
  edge set of a random graph, Random Structures and Algorithms 25~(4) (2004)
  432.

\bibitem{Schrenk2012b}
K.~J. Schrenk, N.~A.~M. Ara\'{u}jo, J.~S. Andrade~Jr, H.~J. Herrmann,
  Fracturing ranked surfaces, Sci. Rep. 2 (2012) 348.

\bibitem{daCosta2010}
R.~A. {da Costa}, S.~N. Dorogovtsev, A.~V. Goltsev, J.~F.~F. Mendes, Explosive
  percolation transition is actually continuous, Phys. Rev. Lett. 105~(25)
  (2010) 255701.

\bibitem{Grassberger2011a}
P.~Grassberger, C.~Christensen, G.~Bizhani, S.-W. Son, M.~Paczuski, Explosive
  percolation is continuous, but with unusual finite size behavior, Phys. Rev.
  Lett. 106~(22) (2011) 225701.

\bibitem{DSouza2010b}
R.~M. D'Souza, M.~Mitzenmacher,
  \href{http://link.aps.org/doi/10.1103/PhysRevLett.104.195702}{Local cluster
  aggregation models of explosive percolation}, Phys. Rev. Lett. 104~(19)
  (2010) 195702.
\newline\urlprefix\url{http://link.aps.org/doi/10.1103/PhysRevLett.104.195702}

\bibitem{bastas2011explosive}
N.~Bastas, K.~Kosmidis, P.~Argyrakis, Explosive site percolation and
  finite-size hysteresis, Phys. Rev. E 84~(6) (2011) 066112.

\bibitem{Choi2011}
Y.~K. Woosik~Choi, Soon-Hyung~Yook, Explosive site percolation with a product
  rule, Phys. Rev. E 84 (2011) 020102.

\bibitem{Tian2012}
T.~Liang, S.~Da-Ning, The nature of explosive percolation phase transition,
  Phys. Lett. A 376 (2012) 286--289.

\bibitem{Fan2012}
F.~Jingfang, M.~Liu, L.~Liangsheng, X.~Chen, Continuous percolation phase
  transition of random networks under a generalized {Achlioptas} process, Phys.
  Rev. E 85 (2012) 061110.

\bibitem{Liu2012}
M.~Liu, J.~Fan, X.~Li, L.S.and~Chen, Continuous percolation phase transition of
  two-dimensional lattice networks under a generalized {Achlioptas} process,
  The European Physical Journal B 85 (2012) 132.

\bibitem{Paris2013}
P.~Giazitzidis, P.~Argyrakis,
  \href{http://link.aps.org/doi/10.1103/PhysRevE.88.024801}{Generalized
  achlioptas process for the delay of criticality in the percolation process},
  Phys. Rev. E 88 (2013) 024801.
\newblock \href {http://dx.doi.org/10.1103/PhysRevE.88.024801}
  {\path{doi:10.1103/PhysRevE.88.024801}}.
\newline\urlprefix\url{http://link.aps.org/doi/10.1103/PhysRevE.88.024801}

\bibitem{Chae2012}
H.~Chae, S.~H. Yook, Y.~Kim, Explosive percolation on the {Bethe} lattice,
  Phys. Rev. E 85~(5 Pt 1) (2012) 051118.

\bibitem{Qian2012}
J.~H. Qian, D.~D. Han, Y.~G. Ma, Criticality and continuity of explosive site
  percolation in random networks, Eur. Phys. Lett. 100 (2012) 48006.

\bibitem{Squires2013}
S.~Squires, K.~Sytwu, D.~Alcala, T.~M. Antonsen, E.~Ott, M.~Girvan, Weakly
  explosive percolation in directed networks, Phys. Rev. E 87 (2013) 052127.

\bibitem{Yi2013}
S.~D. Yi, S.~J. Woo, J.~K. Beom, S.~Seung-{Woo}, Percolation properties of
  growing networks under an {Achlioptas} process, Eur. Phys. Lett. 103 (2013)
  26004.

\bibitem{Hooyberghs2013b}
H.~Hooyberghs, B.~V. Schaeybroeck, J.~O. Indekeu, Degree-dependent network
  growth: From preferential attachment to explosive percolation,
  arXiv:1310.1703v1.

\bibitem{Boettcher2012}
S.~Boettcher, V.~Singh, R.~Ziff, Ordinary percolation with discontinuous
  transitions, Nature Communications 3 (2012) 787.

\bibitem{Angst2012}
S.~Angst, S.~Dahmen, H.~Hinrichsen, A.~Hucht, M.~P. Magiera, Explosive {Ising},
  Journal of Statistical Mechanics: Theory and Experiment (2012) L06002.

\bibitem{Cao2012}
L.~Cao, J.~M. Schwarz, Correlated percolation and tricriticality, Phys. Rev. E
  86~(6 Pt 1) (2012) 061131.

\bibitem{Choi2012}
W.~Choi, S.~H. Yook, Y.~Kim, Bond-site duality and nature of the
  explosive-percolation phase transition on a two-dimensional lattice, Phys.
  Rev. E 86~(5 Pt 1) (2012) 051126.

\bibitem{Lv2012}
J.~P. Lv, X.~Yang, Y.~Deng, Scaling of cluster heterogeneity in the
  two-dimensional {Potts} model, Phys. Rev. E 86~(2 Pt 1) (2012) 022105.

\bibitem{Chi2012}
L.~Chi, X.~Cai, The transition from {E}rd\'{o}s - {R}\'{e}nyi percolation to
  explosive percolation under the partial product rule, International Journal
  of Modern Physics C 23 (2012) 1250083.

\bibitem{Cho2010a}
Y.~S. Cho, S.~W. Kim, J.~D. Noh, B.~Kahng, D.~Kim, Finite-size scaling theory
  for explosive percolation transitions, Phys. Rev. E 82~(4 Pt 1) (2010)
  042102.

\bibitem{Cho2011}
Y.~S. Cho, B.~Kahng, Suppression effect on explosive percolation, Phys. Rev.
  Lett. 107~(27) (2011) 275703.

\bibitem{Lee2011}
H.~K. Lee, B.~J. Kim, H.~Park,
  \href{http://pre.aps.org/abstract/PRE/v84/i2/e020101}{Continuity of the
  explosive percolation transition}, Phys. Rev. E 84~(2) (2011) 020101.
\newline\urlprefix\url{http://pre.aps.org/abstract/PRE/v84/i2/e020101}

\bibitem{Veremyev2012}
A.~Veremyev, V.~Boginski, P.~A. Krokhmal, D.~E. Jeffcoat,
  \href{http://dx.doi.org/10.1371/journal.pone.0051883}{Dense percolation in
  large-scale mean-field random networks is provably "explosive"}, {PLoS} {One}
  7~(12) (2012) e51883.
\newblock \href {http://dx.doi.org/10.1371/journal.pone.0051883}
  {\path{doi:10.1371/journal.pone.0051883}}.
\newline\urlprefix\url{http://dx.doi.org/10.1371/journal.pone.0051883}

\bibitem{Panagiotou2011}
K.~Panagiotou, R.~Sp\"{o}hel, A.~Steger, H.~Thomas, Explosive percolation in
  erd\"{o}s-r\'{e}nyi-like random graph processes, Electronic Notes in Discrete
  Mathematics 38 (2011) 699--704.

\bibitem{Panagiotou2013}
K.~Panagiotou, R.~Sp\"{o}hel, A.~Steger, H.~Thomas, Explosive percolation in
  {Erd\"{o}s}-{R\'{e}nyi}-like random graph processes, Combinatorics,
  Probability and Computing 22 (2013) 133--145.
\newblock \href {http://dx.doi.org/10.1017/S0963548312000442}
  {\path{doi:10.1017/S0963548312000442}}.

\bibitem{Gomez2011}
J.~G\'{o}mez-Garden\~{e}s, S.~C\'{o}mez, A.~Arenas, Y.~Moreno, Explosive
  synchronization transitions in scale-free networks, Phys. Rev. Lett. 106
  (2011) 128701.

\bibitem{Chen2013arxiv}
W.~Chen, X.~Cheng, Z.~Zheng, N.~N. Chung, R.~M. D'Souza, J.~Nagler,
  \href{http://link.aps.org/doi/10.1103/PhysRevE.88.042152}{Unstable
  supercritical discontinuous percolation transitions}, Phys. Rev. E 88 (2013)
  042152.
\newblock \href {http://dx.doi.org/10.1103/PhysRevE.88.042152}
  {\path{doi:10.1103/PhysRevE.88.042152}}.
\newline\urlprefix\url{http://link.aps.org/doi/10.1103/PhysRevE.88.042152}

\bibitem{Chung2013}
N.~N. Chung, L.~Y. Chew, C.~H. Lai, Spectral analysis on explosive percolation,
  Eur. Phys. Lett. 101 (2013) 66003.

\bibitem{Kachhvah2012}
A.~D. Kachhvah, N.~Gupte, Transmission of packets on a hierarchical network:
  statistics and explosive percolation, Phys. Rev. E 86~(2 Pt 2) (2012) 026104.

\bibitem{Cho2011b}
Y.~S. Cho, B.~Kahng, Discontinuous percolation transitions in real physical
  systems, Phys. Rev. E 84 (2011) 050102 (R).

\end{thebibliography}







\end{document}